\documentclass[prb,preprint,showpacs,preprintnumbers,amsmath,amssymb]{revtex4}

\usepackage{graphicx,subfigure}
\usepackage{dcolumn}
\usepackage{bm}
\usepackage{psfrag}
\usepackage{multirow}

\def\be{\begin{equation}}       \def\ee{\end{equation}}
\def\bea{\begin{eqnarray}}      \def\eea{\end{eqnarray}}
\def\half{\frac{1}{2}}
\def\dag{\dagger}
\def\non{\nonumber}
\begin{document}

\title{Spin-wave energy dispersion of a frustrated spin-$\half$ Heisenberg antiferromagnet 
on a stacked square lattice}
\author{Kingshuk Majumdar}
\affiliation{Department of Physics, Grand Valley State University, Allendale, 
Michigan 49401, USA}
\email{majumdak@gvsu.edu}
\date{\today}

\begin{abstract}\label{abstract}
Effects of interlayer coupling and spatial anisotropy on spin-wave excitation spectra of a three-dimensional spatially anisotropic, frustrated spin-$\half$ Heisenberg antiferromagnet (HAFM) is investigated for the two ordered phases using second-order spin-wave expansion. We show that the second-order corrections to the spin-wave energies are significant and find that the energy spectra of the three-dimensional HAFM shares similar qualitative features with the energy spectra of the two-dimensional HAFM on a square lattice. We also discuss the features that can provide experimental measures for the strength of the interlayer coupling, spatial anisotropy parameter, and magnetic frustration. 
\end{abstract}
\pacs{75.10.Jm, 75.40.Mg, 75.50.Ee, 73.43.Nq}

\maketitle
The intriguing properties of many layered antiferromagnets and recently discovered iron based superconductors have created considerable interest in two and three dimensional spin-$\half$ Heisenberg antiferromagnet (HAFM) with competing interactions.\cite{diep} For the last two decades the properties of quantum spin-$\half$ Heisenberg antiferromagnet (HAFM) with nearest neighbor (NN) $J_1$ and next nearest neighbor exchange interactions (NNN) $J_2$ on a square lattice have been studied extensively by various analytical and numerical techniques.~\cite{majumdar10b,
majumdar11a} Earlier studies on the $J_1-J_2$ model have shown that the ground state of the HAFM is antiferromagnetically (AF) ordered at low temperatures. Addition of $J_2$ interactions destabilize the AF order and for a critical value of the frustration parameter 
$\eta=J_2/J_1$ a quantum disordered phase emerges. With increasing values of $J_2$ there is a second quantum phase transition from the disordered phase to a columnar antiferromagnetic (CAF) stripe phase. Using experimental techniques such as nuclear magnetic resonance, magnetization, specific heat, and muon spin rotation measurements, properties of layered magnetic materials Li$_2$VOSiO$_4$, Li$_2$VOGeO$_4$, VOMoO$_4$, and BaCdVO(PO$_4$)$_2$ have been studied.\cite{pavarini08,melzi00,melzi01,bombardi04,carretta02} These studies have shown that these compounds have significant couplings between NN and NNN neighbors. Moreover, for Li$_2$VOSiO$_4$, a layered material that can be described by a square lattice $J_1-J_2$ model with large $J_2$ the interlayer  coupling $J_\perp/J_1 \sim 0.07$ is not negligible.\cite{rosner02} Due to a finite interlayer magnetic coupling $J_\perp$, these experimental systems are quasi 2D.

Experimentally one of the most direct ways to probe the magnetic excitation spectra is with inelastic neutron scattering (INS). For example, INS experiments on $S=1/2$ 2D antiferromagnets Sr$_2$Cu$_3$O$_4$Cl$_2$, Cu(DCOO)$_2\cdot$4D$_2$O (CFTD), and La$_2$CuO$_4$ (LCO) have revealed that the magnon energy at the wave-vector $(\pi,0)$ is depressed by 7\% relative to the energy at the wave-vector $(\pi/2,\pi/2)$.\cite{ronnow01,chris04,chris07} Another example is the recently discovered iron pnictide superconductors.\cite{kamihara08} In the parent phases of these materials magnetic excitations have shown to play an important role in the superconducting state.\cite{kamihara08,cruz08,klaus08,Dong08,apple10,singh09,uhrig09,yao08,yao10} INS measurements on CaFe$_2$As$_2$ and SrFe$_2$As$_2$ have revealed a very large difference between $J_1$ and $J_1^\prime$.\cite{zhao09,han09,schmidt10}

Theoretically the spin-wave spectra and the low-temperature magnetic phase diagram have been obtained for the spatially anisotropic $J_1-J_1^\prime-J_2$ HAFM model on a square lattice with NN exchanges $J_1$ along the $x$ axis, $J_1^\prime$ along the $y$ axis, and NNN interactions $J_2$ along the diagonals in the $xy$ plane.\cite{majumdar10b} Recent experiments on iron-based superconductors such as undoped iron oxypnictides reveal that the electronic couplings are more three dimensional than in the cuprate superconductors.\cite{zhao08,ewings08,yuan09} With decrease in temperature most undoped iron-pnictide superconductors show a structural transition from a tetragonal paramagnetic phase to a orthorhombic phase. In the 122 materials a three dimensional (3D) long-range
antiferromagnetic order develops simultaneously. The magnetic phase diagram of the 3D $J_1-J_1^\prime-J_2-J_\perp$ HAFM model with interlayer coupling $J_\perp$ has also been recently investigated.\cite{majumdar11a,schmidt10,holt10}. The present work is a sequel of  Ref.~\onlinecite{majumdar11a} by the author. Here we study the spin-wave energy dispersion of the two ordered phases of the model taking into account the first ($1/S$) and second-order ($1/S^2$) corrections 
to the Hamiltonian. The details of the model and the derivations along with many references can be found in Ref.~\onlinecite{majumdar11a}.

The Heisenberg Hamiltonian for a spatially anisotropic, frustrated spin-$\half$ HAFM on a cubic lattice with four exchange AF interactions between spins: $J_1$ along the $x$ (row) direction, $J_1^\prime$ along the $y$ (column) direction, $J_2$ along the diagonals in the $xy$ plane, and with interlayer coupling $J_\perp$ is described by 
\be
H = \half \sum_{i, \ell} \Big[J_1{\bf S}_{i,\ell} \cdot {\bf S}_{i+\delta_x,\ell}
+ J_1^\prime {\bf S}_{i,\ell} \cdot {\bf S}_{i+\delta_y,\ell} 
+ J_2 {\bf S}_{i,\ell} \cdot {\bf S}_{i+\delta_x + \delta_y,\ell}\Big]
+\half J_\perp \sum_{i,\ell}{\bf S}_{i,\ell} \cdot {\bf S}_{i,\ell+1}.
\label{hamiltonian}
\ee
$\ell$ labels the layers, $i$ runs over $N_L$ lattice sites, $i+\delta_x$ ($\delta_x =\pm 1$) 
and $i+\delta_y$ ($\delta_y =\pm 1$) are the 
NN interactions to the $i$-th site along the row and the column direction. The third
term represents the NNN interaction, which are along the
diagonals in the $xy$ plane and the last term is for the NN coupling between the layers. 
$\zeta = J_1^\prime/J_1$ measures the directional
anisotropy, $\eta=J_2/J_1$ is the magnetic frustration between the 
NN (row direction) and NNN spins, and $\delta=J_\perp/J_1$ is the interlayer coupling 
parameter. 
At zero temperature the classical ground states for the anisotropic model are the 
Ne\'{e}l or the AF ($\pi,\pi, \pi$) state and the columnar antiferromagnetic (CAF) ($\pi,0,\pi$)
state (the classical transition occurs at $\eta_c=\zeta/2$).

For the AF ordered phase, NN interactions are between A and B sublattices and NNN interactions are between A-A and B-B sublattices. The Hamiltonian in Eq.~\ref{hamiltonian}
takes the form:
\bea
H &=& J_1 \sum_{i,\ell}{\bf S}_{i,\ell}^{\rm A} \cdot {\bf S}_{i+\delta_x,\ell}^{\rm B}
+ J_1^\prime \sum_{i,\ell}{\bf S}_{i,\ell}^{\rm A} \cdot {\bf S}_{i+\delta_y,\ell}^{\rm B} 
+ \half J_2 \sum_{i,\ell}\Big[ {\bf S}_{i,\ell}^{\rm A} \cdot {\bf S}_{i+\delta_x + \delta_y,\ell}^{\rm A}
+ {\bf S}_{i,\ell}^{\rm B} \cdot {\bf S}_{i+\delta_x + \delta_y,\ell}^{\rm B}\Big]\nonumber \\
&+& J_\perp \sum_{i,\ell}{\bf S}_{i,\ell}^{\rm A} \cdot {\bf S}_{i,\ell+1}^{\rm B}.
\label{ham-AF}
\eea
This Hamiltonian is mapped to bosonic creation 
and annihilation operators $a^\dag, a$ and $b^\dag, b$ using
Holstein-Primakoff transformations keeping only terms up to the order of $1/S^2$. Furthermore, the Hamiltonian is diagonalized in the momentum space using the  Bogoliubov transformations. 
In powers of $1/S$ the Hamiltonian can be written as\cite{majumdar11a}
\be
H = -\half N J_1 S^2 z(1+\zeta)\Big[1 - \frac {2\eta}{1+\zeta}\Big] + H_0 + H_1+H_2 + \ldots,
\ee 
where 
\bea
H_0 &=& J_1Sz(1+\zeta)\sum_{\bf k} \kappa_{\bf k}\left(\epsilon_{\bf k} -1\right)
+J_1Sz(1+\zeta)\sum_{\bf k}\kappa_{\bf k} \epsilon_{\bf k}
\left( \alpha^\dag_{\bf k}\alpha_{\bf k}+\beta^\dag_{\bf k}\beta_{\bf k}\right),
\label{H0term} \\
H_1 &=& \frac {J_1Sz(1+\zeta)}{2S}\sum_{\bf k}
\Big[ A_{\bf k}\left(\alpha^\dag_{\bf k}\alpha_{\bf k}+
\beta^\dag_{\bf k}\beta_{\bf k}\right) 
+ B_{\bf k}\left(\alpha^\dag_{\bf k}\beta_{-\bf k}^\dag+
\beta_{-\bf k}\alpha_{\bf k}\right)\Big] \non \\
&-& \frac {J_1Sz(1+\zeta)}{2SN}\sum_{1234}
\delta_{\bf G}(1+2-3-4)l_1l_2l_3l_4\Big[\alpha_1^\dag \alpha_2^\dag \alpha_3 \alpha_4
V_{1234}^{(1)} +\beta^\dag_{-3}\beta^\dag_{-4}\beta_{-1}\beta_{-2}V_{1234}^{(2)} \non \\
&+&4\alpha_1^\dag \beta_{-4}^\dag \beta_{-2}\alpha_3 V_{1234}^{(3)} +\Big\{
2\alpha_1^\dag \beta_{-2}\alpha_3 \alpha_4V_{1234}^{(4)} +2\beta_{-4}^\dag \beta_{-1}
\beta_{-2}\alpha_3 V_{1234}^{(5)} + \alpha_1^\dag \alpha_2^\dag \beta_{-3}^\dag 
\beta_{-4}^\dag V_{1234}^{(6)} \non \\
&+& h.c.\Big\}\Big],
\label{H1term} \\
H_2 &=& \frac {J_1Sz(1+\zeta)}{(2S)^2} \sum_{\bf k} 
\Big[ C_{1{\bf k}}\left(\alpha^\dag_{\bf k}\alpha_{\bf k}+\beta^\dag_{\bf k}\beta_{\bf k}
\right)+C_{2{\bf k}}\left(\alpha^\dag_{\bf k}\beta_{-\bf k}^\dag+
\beta_{-\bf k}\alpha_{\bf k}\right)+\ldots \Big].
\label{H2term}
\eea
In the above equation,
\bea
\epsilon_{\bf k} &=& (1-\gamma_{\bf k}^2)^{1/2},  \non \\
\gamma_{1{\bf k}}&=&[\cos (k_x)+\zeta \cos (k_y)+\delta \cos(k_z)]/(1+\zeta),
\;  \gamma_{2{\bf k}} = \cos (k_x)\cos(k_y), \label{defs-AF} \\
\gamma_{\bf k} &=& \gamma_{1{\bf k}}/\kappa_{\bf k},\; \kappa_{\bf k} = 1- \frac {2\eta}{1+\zeta} (1-\gamma_{2{\bf k}})+\frac{\delta}{1+\zeta}.\non
\eea

The first term in $H_0$ is the zero-point energy and the second term represents the 
excitation energy of the magnons within linear spin-wave theory (LSWT). The part $H_1$ and $H_2$ correspond to $1/S$ and 
$1/S^2$ corrections to the Hamiltonian.

The spin-wave energy ${\tilde E_{\bf k}^{\rm AF}}$ for magnon excitations, measured 
in units of $J_1Sz(1+\zeta)$ up to second order in $1/S$ for the AF-phase is given as\cite{majumdar10b}
\be
{\tilde E_{\bf k}^{\rm AF}} = E_{\bf k} + \frac 1{(2S)} A_{\bf k}+
\frac 1{(2S)^2}\Big[\Sigma^{(2)}_{\alpha \alpha}({\bf k},E_{\bf k})-
\frac {B_{\bf k}^2}{2E_{\bf k}} \Big],
\label{energyEk-AF}
\ee
where the second-order self-energy is given by 
\bea
\Sigma_{\alpha \alpha}^{(2)}({\bf k},\omega) &=& \Sigma_{\beta \beta}^{(2)}(-{\bf k},-\omega)
=C_{1{\bf k}} 
+ \Big(\frac {2}{N} \Big)^2\sum_{{\bf pq}}2l_{\bf k}^2l_{\bf p}^2l_{\bf q}^2l_{\bf k+p-q}^2 \non \\
&\times& \Big[\frac {|V^{(4)}_{\bf k,p,q,[k+p-q]}|^2}{\omega -E_{\bf p}
-E_{\bf q}-E_{\bf k+p-q}+i0^+} - \frac {|V^{(6)}_{\bf k,p,q,[k+p-q]}|^2}{\omega +E_{\bf p}
+E_{\bf q}+E_{\bf k+p-q}+i0^-}  \Big].\label{sigma1}
\eea
The coefficients $\ell_{\bf k}, A_{\bf k}, B_{\bf k}, C_{1{\bf k}}$ are given in Ref.~\onlinecite{majumdar11a}.

On the other hand, the Hamiltonian describing the CAF phase is
\bea
H &=& J_1 \sum_{i,\ell}{\bf S}_{i,\ell}^{\rm A} \cdot {\bf S}_{i+\delta_x,\ell}^{\rm B}
+ \half J_1^\prime\sum_{i,\ell}\Big[ {\bf S}_{i,\ell}^{\rm A} \cdot {\bf S}_{i+ \delta_y,\ell}^{\rm A}
+ {\bf S}_{i,\ell}^{\rm B} \cdot {\bf S}_{i+ \delta_y,\ell}^{\rm B}\Big]
+ J_2 \sum_{i,\ell}{\bf S}_{i,\ell}^{\rm A} \cdot {\bf S}_{i+\delta_x +\delta_y,\ell}^{\rm B} \non \\
&+& J_\perp \sum_{i,\ell}{\bf S}_{i,\ell}^{\rm A} \cdot {\bf S}_{i,\ell+1}^{\rm B}.
\label{ham-CAF}
\eea
The quasiparticle energy ${\tilde E_{\bf k}^{\rm CAF}}$, measured 
in units of $J_1Sz(1+2\eta)$ up to second order in $1/S$ for the CAF phase is 
\be
{\tilde E_{\bf k}^{\rm CAF}} = E_{\bf k}^\prime + \frac 1{2S} A_{\bf k}^\prime+
\frac 1{(2S)^2}\Big[\Sigma^{\prime (2)}_{\alpha \alpha}({\bf k},E_{\bf k}^\prime)-
\frac {B_{\bf k}^{\prime 2}}{2E_{\bf k}^\prime} \Big].
\label{energyEk-CAF}
\ee
The structure factors $\gamma_{1{\bf k}}^\prime,\; 
\gamma_{2{\bf k}}^\prime$ in this case are
\bea
\gamma^\prime_{1{\bf k}}&=&\big[\cos (k_x)(1+2\eta \cos (k_y))+\delta \cos (k_z)\big]/(1+2\eta), 
\; \gamma^\prime_{2{\bf k}} = \cos(k_y),  \non\\
\gamma^\prime_{\bf k} &=& \gamma^\prime_{1{\bf k}}/\kappa^\prime_{\bf k},\;
\kappa_{\bf k}^\prime = 1- \frac {\zeta}{1+2\eta} (1-\gamma^\prime_{2{\bf k}})+\frac{\delta}{1+2\eta},\\
\epsilon_{\bf k}^\prime &=& [1-\gamma_{\bf k}^{\prime 2}]^{1/2}. \non  
\eea
The different coefficients can be found in Ref.~\onlinecite{majumdar11a}.

Spin-wave energy dispersions for the
two ordered phases with different values of $\zeta$,
$\eta$, and $\delta$ are obtained numerically from Eqs.~\ref{energyEk-AF} and \ref{energyEk-CAF}. 
Especially to obtain the second order correction terms we sum up contributions from $N_L^3$ points of ${\bf p}$ and $N_L^3$ points of ${\bf q}$ in the first BZ. For our results we use $N_L=18$ lattice sites.
\begin{figure}[httb]
\includegraphics[width=3.0in]{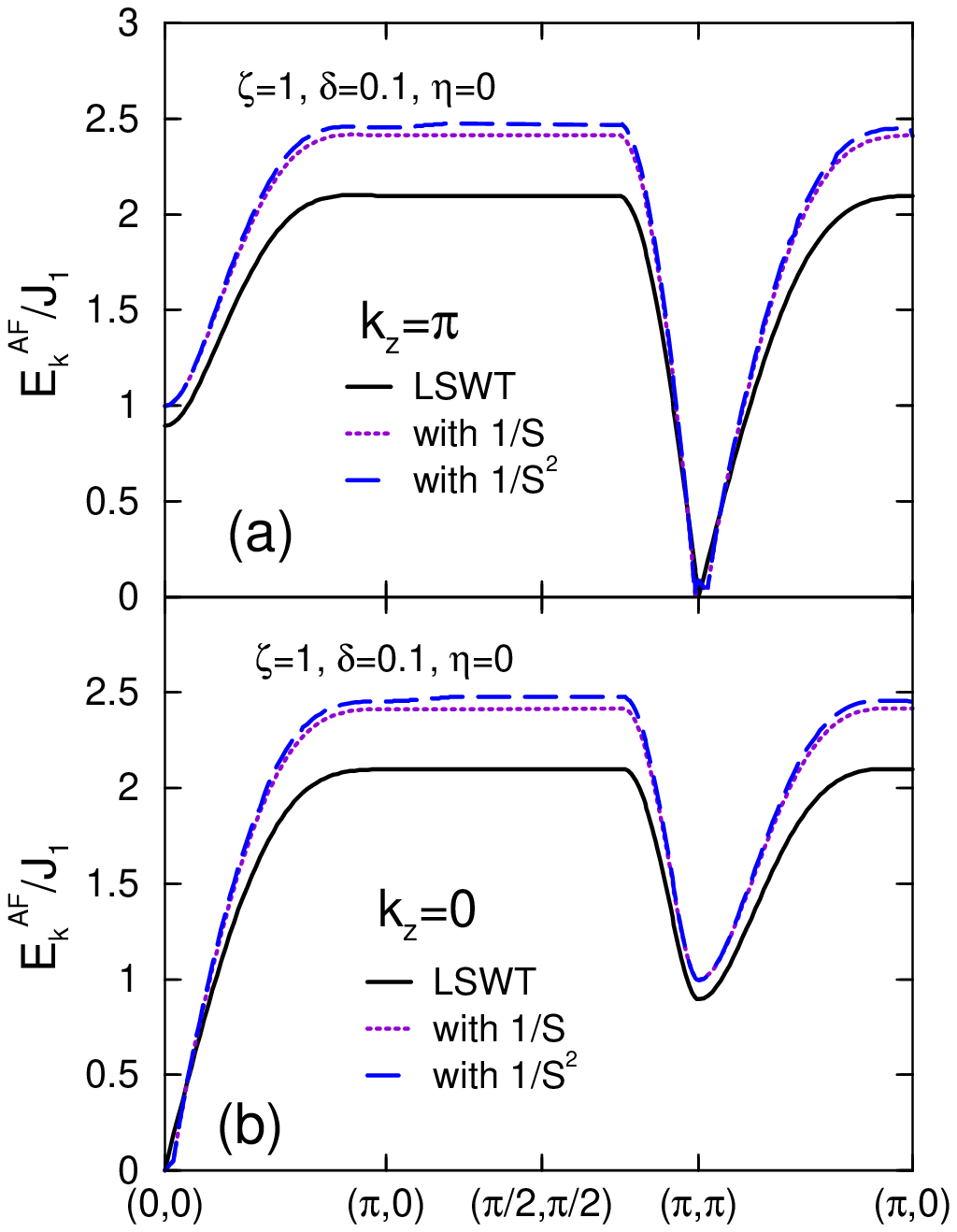}
\includegraphics[width=3.0in]{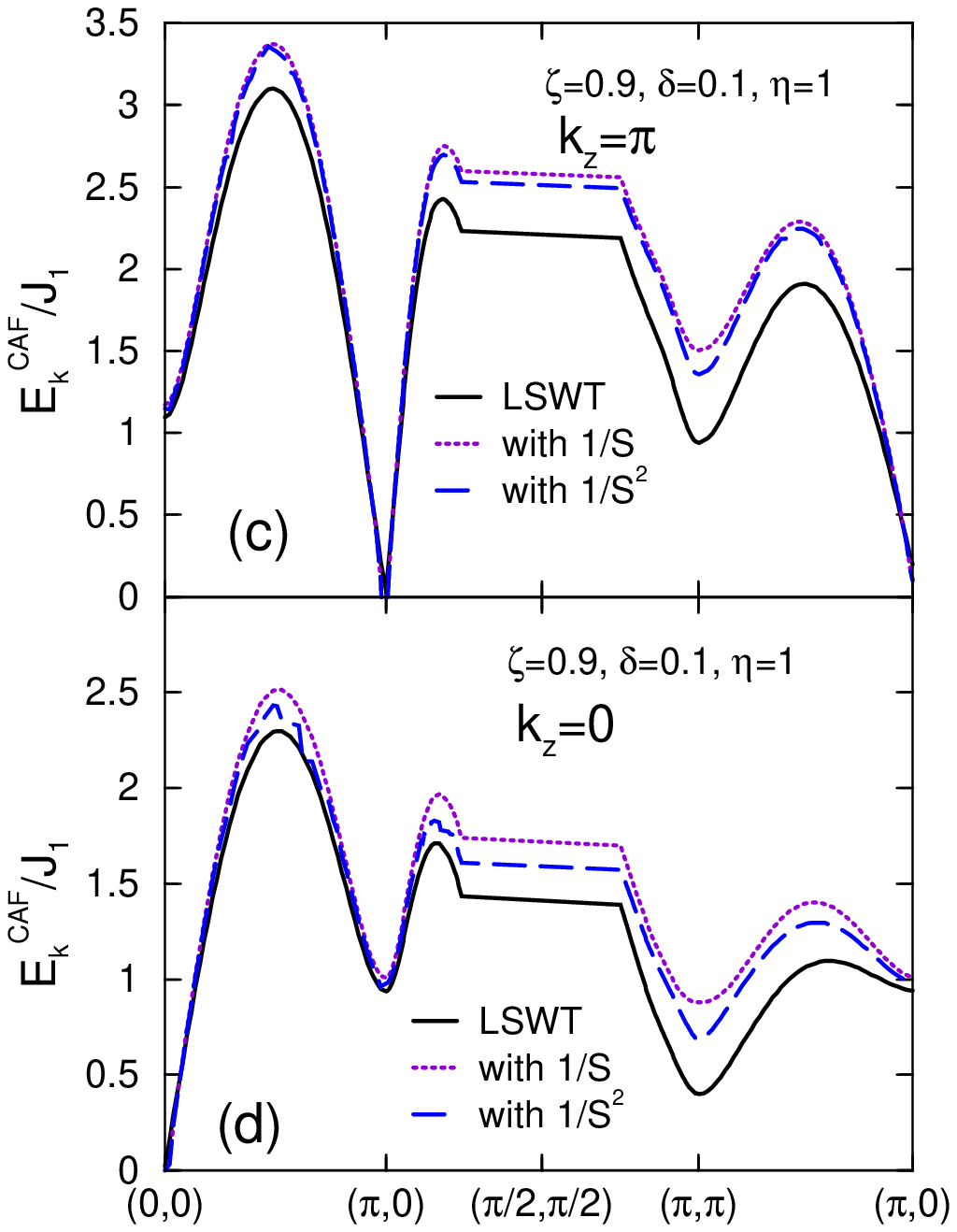}
\caption{\label{fig:EnergyComp} (Color online) Spin-wave energies $E^{\rm AF}_{\bf k}/J_1$ for the AF phase (panels a and b)  with $\zeta=1, \delta=0.1$, and $\eta=0$ and 
for the CAF phase (panels c and d) with $\zeta=0.9, \delta=0.1$, and $\eta=1$ obtained
from LSWT (solid lines), with first-order corrections (dotted lines), and with second-order
corrections (long dashed lines). Momenta $(k_x,k_y)$ are varied from (0,0) to $(\pi, \pi)$ in the first BZ keeping $k_z=\pi$ (panels a and c) and $k_z=0$ (panels b and d). The local energy minima are at $(\pi, \pi,\pi)$ for the AF phase and at $(\pi,0,\pi)$ for the CAF phase. For both the phases the first and second order corrections significantly increase the spin-wave energy from the results obtained using LSWT. For the CAF phase $1/S^2$ corrections lower the spin-wave energy from the first-order corrections.}
\end{figure}

Figure~\ref{fig:EnergyComp} shows the spin-wave energies $E^{\rm AF}_{\bf k}/J_1$ for the AF phase (panels a and b) with $\zeta=1, \delta=0.1, \eta=0$ and for the CAF phase (panels c and d) with $\zeta=0.9, \delta=0.1, \eta=1$ obtained from LSWT (solid lines), with first-order corrections (dotted lines), and with second-order corrections (long dashed lines). Momenta $(k_x,k_y)$ are varied from (0,0) to $(\pi, \pi)$ in the first BZ with $k_z$ fixed at $\pi$ or 0 as $(0,0,\pi) \rightarrow (\pi,0,\pi) \rightarrow (\pi/2,\pi/2,\pi) \rightarrow (\pi,\pi,\pi) 
\rightarrow (\pi,0,\pi)$ (panels a and c) and $(0,0,0) \rightarrow (\pi,0,0) \rightarrow (\pi/2,\pi/2,0) \rightarrow (\pi,\pi,0) 
\rightarrow (\pi,0,0)$ (panels b and d). Panels (a) and (b) have Goldstone modes at the wave-vectors ${\bf k}=0$ and ${\bf k}=( \pi,\pi, \pi)$
whereas for panels (c) and (d) the Goldstone modes are at ${\bf k}=0$ and ${\bf k}=(\pi,0, \pi)$. The local energy minima are at $(\pi,\pi,\pi)$ for the AF phase and at $(\pi,0,\pi)$ for the CAF phase. For both the phases the first and second order corrections significantly increase the spin-wave energy from that obtained with LSWT. However, for the CAF phase $1/S^2$ corrections lower the spin-wave energy from the first-order corrections.

\begin{figure}[httb]
\includegraphics[width=3.0in]{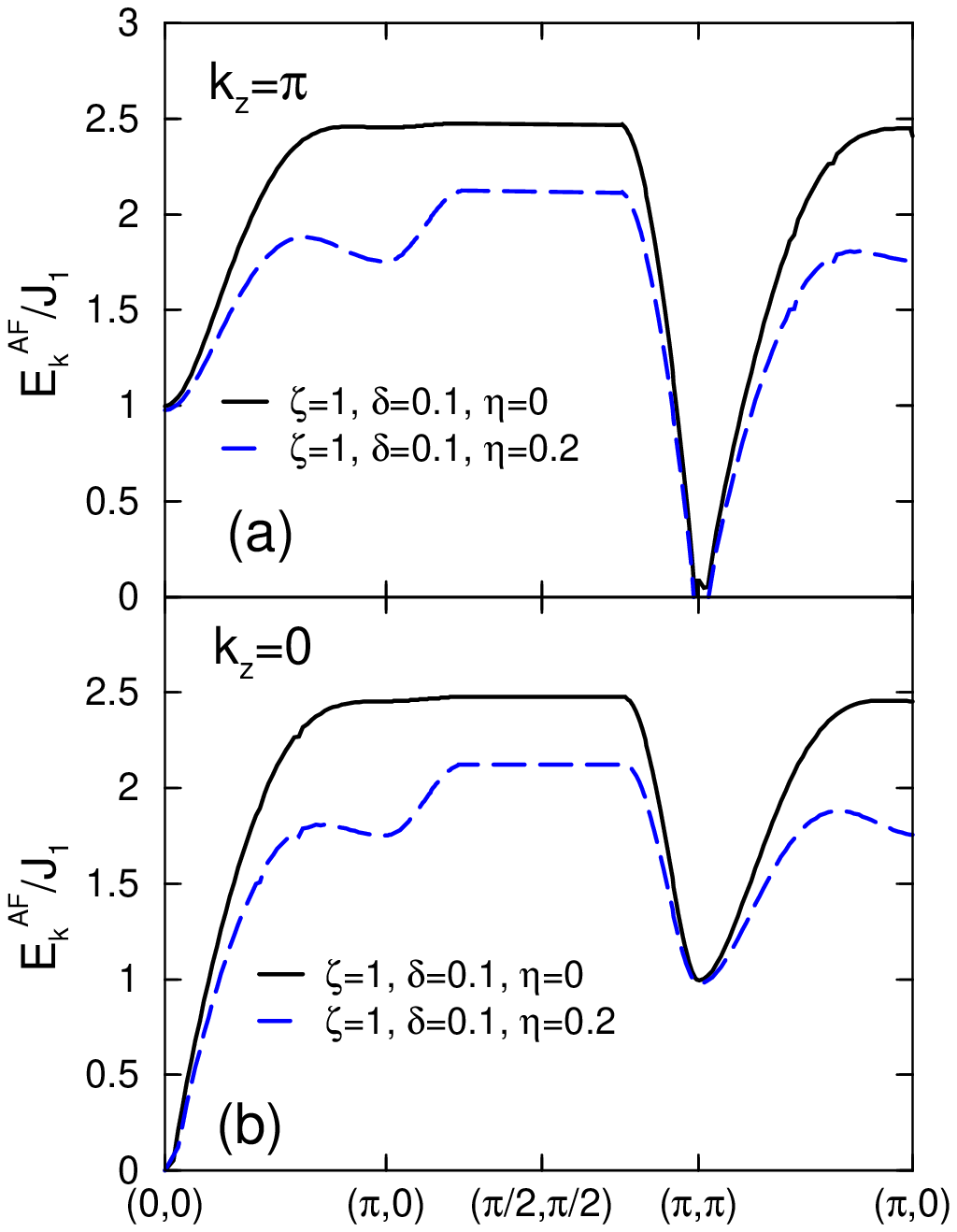}
\includegraphics[width=3.0in]{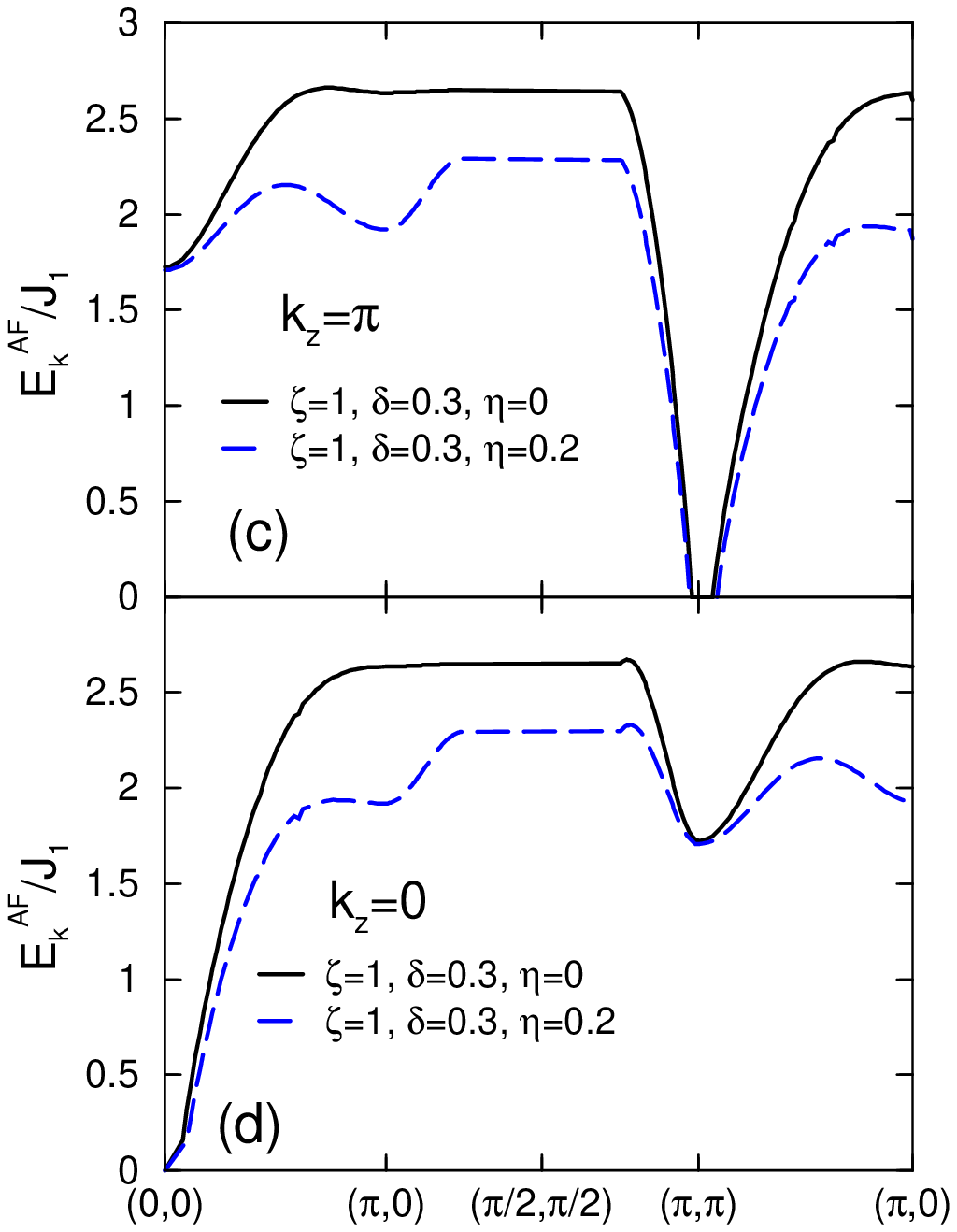}
\caption{\label{fig:EnergyAF} (Color online) Plot of spin-wave energy dispersion (with second-order corrections) $E^{\rm AF}_{\bf k}/J_1$ for the AF phase with $\zeta=1$, $\delta=0.1$ (panels a and b), 0.3 (panels c and d), $\eta=0$, and $0.2$. $k_z=\pi$ for panels a and c and $k_z=0$ for panels b and d. Minimum of energy occurs at $(\pi,\pi,\pi)$ for the AF phase (the narrow $E^{\rm AF}_{\bf k}=0$ window near $(\pi,\pi,\pi)$ is due to numerical inaccuracy). With increase in NNN frustration the dip in the spin-wave energies at $(\pi,0,\pi)$ and $(\pi,0,0)$ increases. This can provide an experimental measure of the strength of NNN frustration. The strength of the interlayer coupling $\delta$ an be measured from the spin-wave energy at $(0,0,\pi)$ (see text).}
\end{figure}

The spin-wave energies for AF phase $E^{\rm AF}_{\bf k}/J_1$ is plotted in Fig.~\ref{fig:EnergyAF}  for different values of $\zeta, \eta, \delta$ in the BZ. The dispersion along the 
$(\pi/2, \pi/2,\pi) - (\pi,0,\pi)$ and $(\pi/2, \pi/2,0) - (\pi,0,0)$ are flat within LSWT and $1/S$
correction (see Fig.~\ref{fig:EnergyComp} for example). The second order corrections make the magnon
energies at $(\pi,0,\pi)$ or $(\pi,0,0)$ smaller than the energies at $(\pi/2,\pi/2,\pi)$
or $(\pi/2,\pi/2,0)$. Also we find that with increase in NNN frustration the dip in the spin-wave energies at $(\pi,0,\pi)$ and $(\pi,0,0)$ increases. This can provide an experimental measure of the strength of the NNN frustration. These features are qualitatively similar to the recently obtained 
spin-wave energy dispersion for the frustrated HAFM on the spatially anisotropic 
square lattice.\cite{majumdar10b}

\begin{figure}[httb]
\includegraphics[width=3.0in]{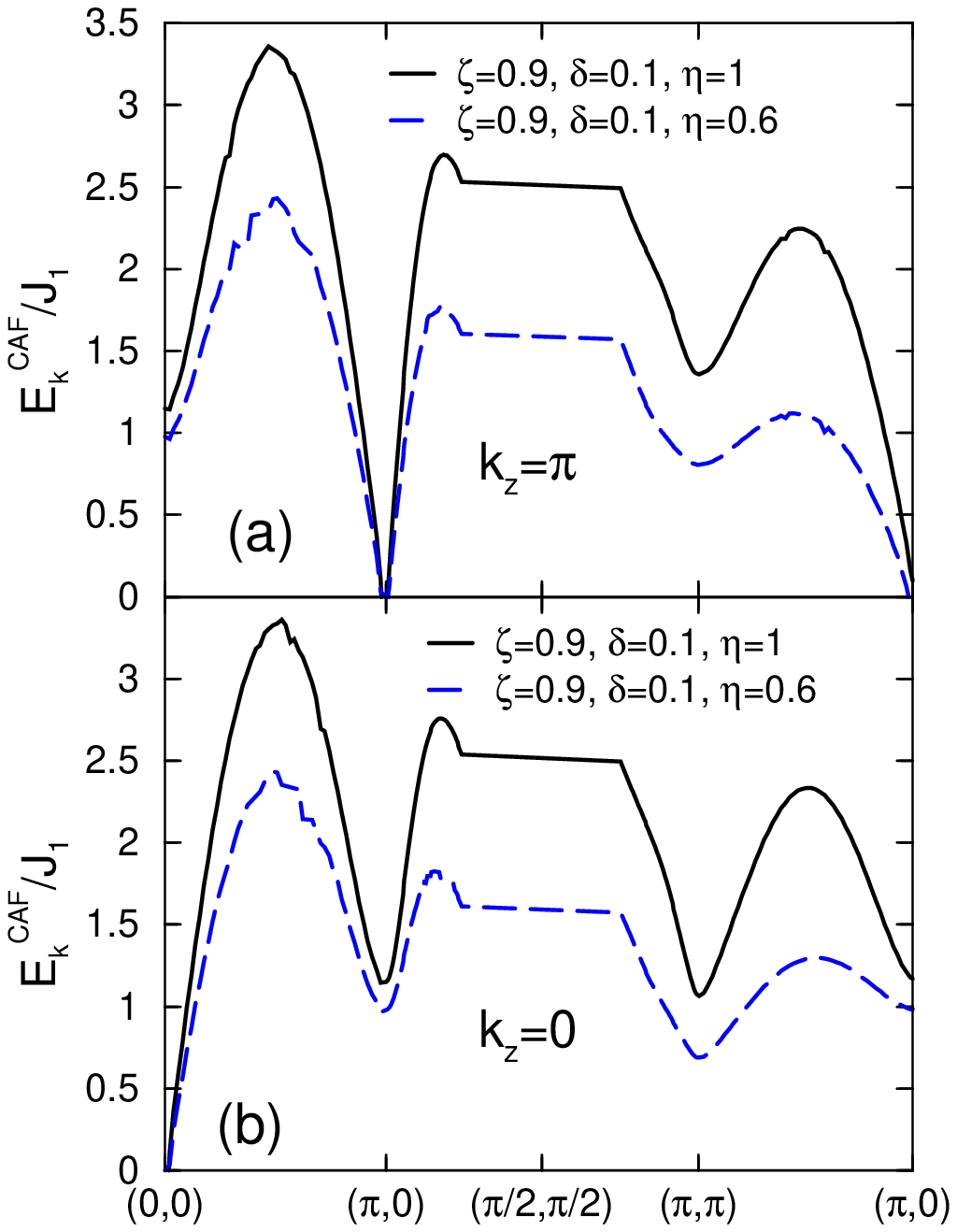}
\includegraphics[width=3.0in]{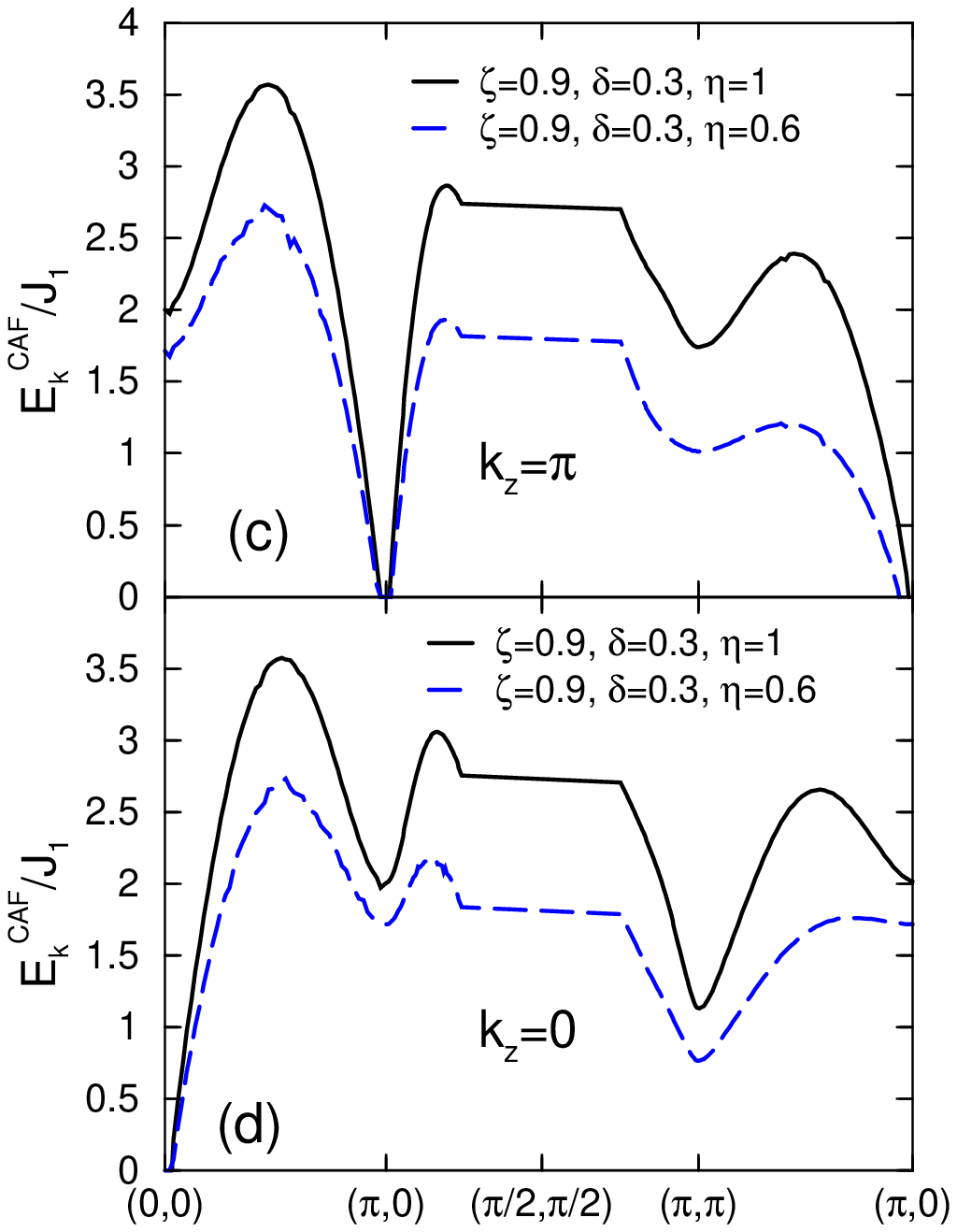}
\caption{\label{fig:EnergyCAF} (Color online) Spin-wave energy dispersion (with second-order corrections) $E^{\rm CAF}_{\bf k}/J_1$ for the CAF phase with $\zeta=0.9$, $\eta=1, 0.6$, and $\delta=0.1$ (panels a and b), 0.3 (panels c and d). Minimum of energy occurs at $(\pi,0,\pi)$ for the CAF phase. The energy gap at the wave-vector $(0,0,\pi)$ increases with increase in $\delta$ (see panels a and c)}
\end{figure}

Figure \ref{fig:EnergyCAF} displays the energy dispersion for the CAF phase with $\zeta=0.9$, $\delta=0.1$ (panels a and b), 0.3 (panels c and d), $\eta=1$ and 0.6. 
We find three energy peaks, the maxima being at $(\pi/2,0,\pi)$ and $(\pi/2,0,0)$. 
With decrease in frustration the system approaches the phase transition 
point, so the spin-wave energy diminishes. 

\begin{figure}[httb]
\includegraphics[width=3.0in]{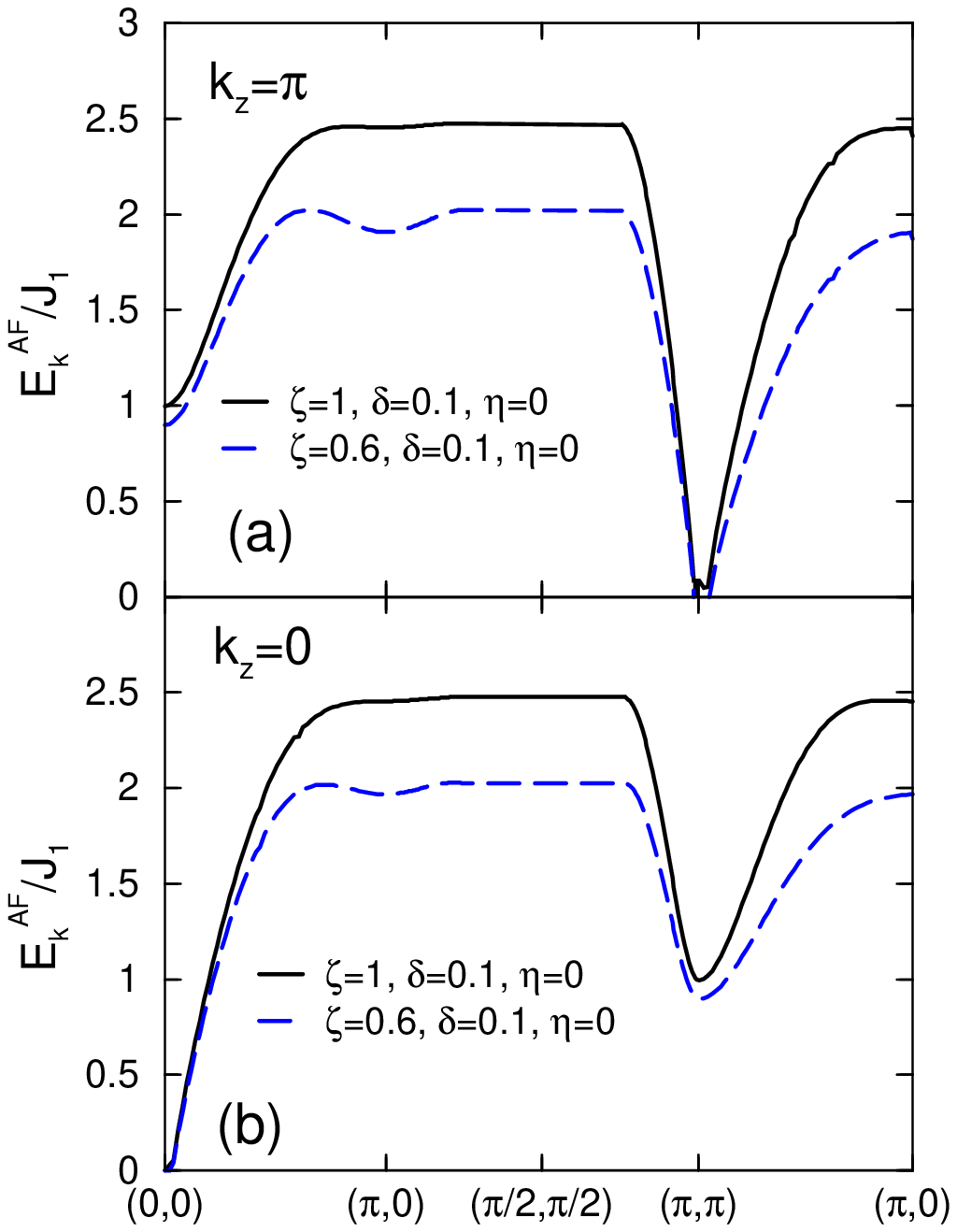}
\includegraphics[width=3.0in]{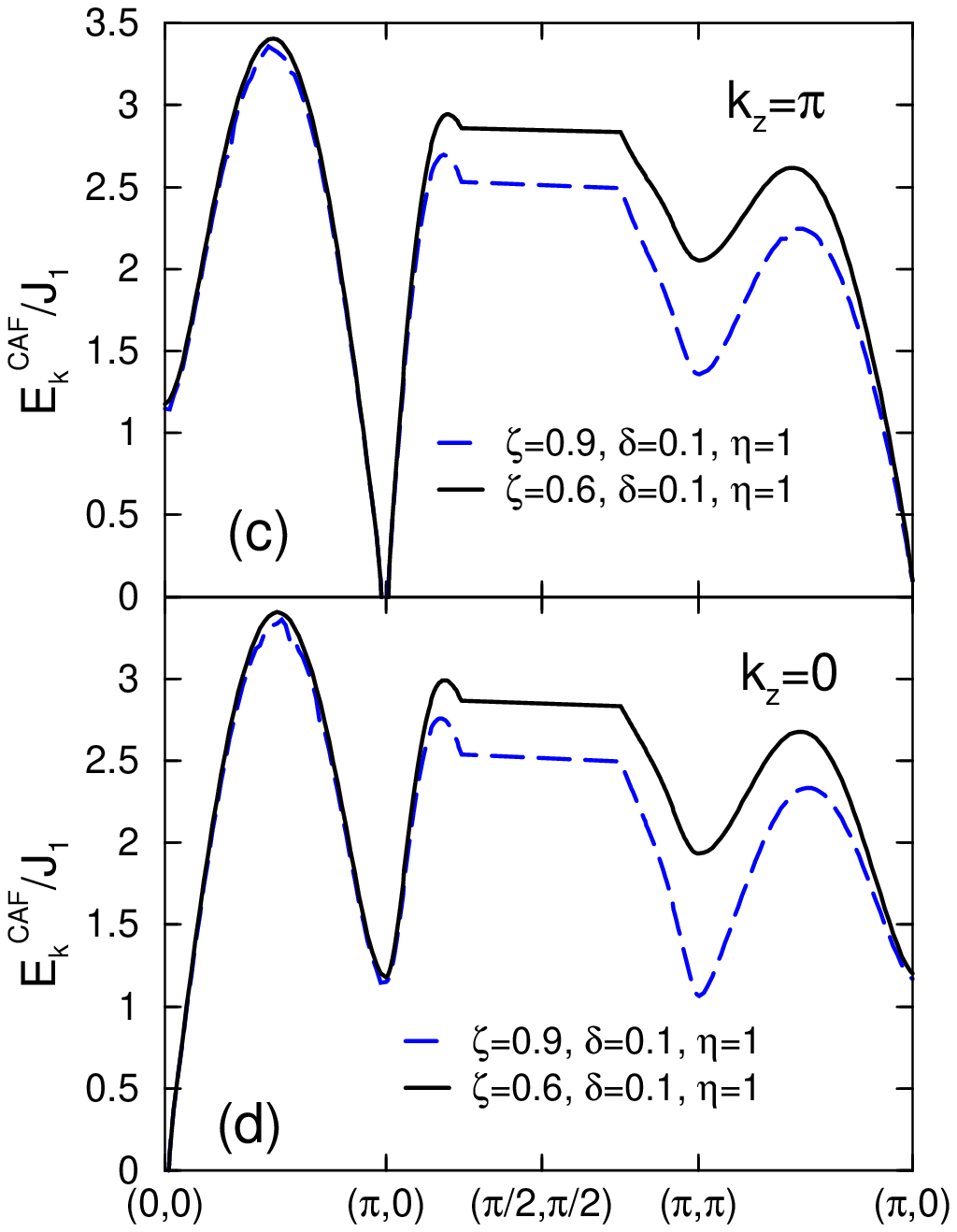}
\caption{\label{fig:EnergyAF-CAF} (Color online) Effects of spatial anisotropy $\zeta$ 
on the energy dispersion for the AF and the CAF phases are shown. For the AF phase with increase
in directional anisotropy (i.e. with lowering values of $\zeta$) the spin-wave energy decreases.
The dip in the energy at $(\pi,0,0)$ or $(\pi,0,\pi)$ can provide an experimental measure of 
$\zeta$. For the CAF phase the trend is opposite, i.e. $E^{\rm CAF}_{\bf k}/J_1$ increases as 
$\zeta$ is lowered from 1 to 0.6. $E^{\rm CAF}(0,0,\pi)/J_1$ is independent of $\zeta$ at the wave-vector $(0,0,\pi)$ (panel c).
}
\end{figure}

For $\delta=0$, we have a Goldstone mode at the wave-vector ${\bf k}=(0,0,\pi)$ for both the AF and CAF phases. But presence of a finite interlayer coupling $\delta$ opens up a gap at these wave-vectors (see Figs.~\ref{fig:EnergyAF} and \ref{fig:EnergyCAF} (panels a and c)). This gap increases with increase in $\delta$. We find that the spin-wave energy $E^{\rm AF}_{\bf k}/J_1$ for the AF-phase obtained from LSWT (using Eq.~\ref{energyEk-AF}) at the wave-vector $(0,0,\pi)$ in the BZ is $2\sqrt{\delta(1+\zeta)}/[1+\delta/(1+\zeta)]$, which is independent of frustration $\eta$ 
(see Fig.~\ref{fig:EnergyAF}a). As an example, within LSWT for $\zeta=1$ we find 
$E^{\rm AF}(0,0,\pi)/J_1 \approx 0.85$ with $\delta=0.1$ (Fig.~\ref{fig:EnergyComp}a). On the other hand, for the CAF phase we find the energy $E^{\rm CAF}(0,0,\pi)/J_1$ (from LSWT) to be $2\sqrt{\delta(1+2\eta)}/[1+\delta/(1+2\eta)]$, which is independent of the anisotropy parameter $\zeta$ (see Fig.~\ref{fig:EnergyAF-CAF}c). In this case we find for $\eta=1$, 
$E^{\rm CAF}(0,0,\pi)/J_1 \approx 1.1$ with $\delta=0.1$ (see Fig.~\ref{fig:EnergyComp}c). However, first and second order corrections slightly increase these LSWT predictions. The strength of the interlayer coupling $\delta$ can be measured experimentally from the energy at this ${\bf k}=(0,0,\pi)$ value.

In Fig.~\ref{fig:EnergyAF-CAF} we show the effects of spatial anisotropy on the two ordered phases of the model. For the AF phase with increase in directional anisotropy (i.e., with lowering values of $\zeta$) the spin-wave energy decreases. The dip in the energy at $(\pi,0,0)$ or $(\pi,0,\pi)$ can provide an experimental measure of $\zeta$. For the CAF phase the trend is opposite, i.e. $E^{\rm CAF}_{\bf k}/J_1$ increases as $\zeta$ is lowered from 1 to 0.6. From panel c we find that at the wave-vector $(0,0,\pi)$, $E^{\rm CAF}(0,0,\pi)/J_1$ is independent of $\zeta$. This is similar to 
$E^{\rm AF}(0,0,\pi)/J_1$ being independent of $\eta$ (see panels a and c of Fig.\ref{fig:EnergyAF})

The present work is a sequel of an earlier theoretical paper where the author studied the magnetic phase diagram of a spatially anisotropic, frustrated spin-$\half$ HAFM on a 3D lattice with interlayer coupling.\cite{majumdar11a} In the present work, we have studied the effects of interlayer coupling and spatial anisotropy on the spin-wave energies of the two long-range ordered phases (antiferromagnetic Ne\'{e}l and antiferromagnetic columnar stripe). We have shown that the second-order corrections to the energy spectra are significant. Also we have found that our obtained energy spectra for the three-dimensional HAFM model (with interlayer coupling) shares similar qualitative features with the energy spectra for the two-dimensional square lattice. Finally, we have provided a few key features in the energy spectra that can be measured experimentally (e.g., with neutron scattering experiments). These measurements can provide us information on the strength of the interlayer coupling, spatial anisotropy parameter, and magnetic frustration. 

{\em Acknowledgments:}
This project acknowledges the use of the Cornell Center for Advanced Computing's ``MATLAB on the TeraGrid'' experimental computing resource funded by NSF grant 0844032 in partnership with Purdue University, Dell, The MathWorks, and Microsoft.

\bibliography{Aniso3DEnergy}

\end{document}